# The quantum metrology triangle and the re-definition of the SI ampere and kilogram; Analysis of a reduced set of observational equations

Martin J T Milton, Jonathan M Williams and Alistair B Forbes.
National Physical Laboratory, Hampton Road, Teddington,
Middlesex, TW11 0LW, UK.

**Short Title** The SI and the fundamental constants

**PACS** 06.20.fa; 06.20.Jr; 73.43.Cd; 74.50

## Abstract

We have developed a set of seven observational equations that include all of the physics necessary to relate the most important of the fundamental constants to the definitions of the SI kilogram and ampere. We have used these to determine the influence of alternative definitions being considered for the SI kilogram and ampere on the uncertainty of three of the fundamental constants ($h$, $e$ and $m_u$). We have also reviewed the experimental evidence for the exactness of the quantum metrology triangle resulting from experiments combining the quantum Hall effect, the Josephson effects and single-electron tunnelling.

## Introduction

There is much debate about the benefits of revising the definitions of several of the base units of the SI [1] so that they are defined in terms of fixed values of certain fundamental constants [2,3,4] whilst the underlying principles and other more pragmatic considerations are central to this debate [5], the final outcome will be strongly influenced by quantitative arguments about the uncertainties of both the disseminated units and the fundamental constants most closely associated with them. In order to evaluate the merits of alternative scenarios for the re-definition of the ampere and the kilogram, we have performed least-squares adjustments of a reduced set of observational equations linking a set of the fundamental constants and quantified their impact on the uncertainties of the Planck constant ($h$), the elementary charge ($e$) and the atomic mass constant ($m_u$). In particular, we have been able to compare the alternatives of fixing either $h$ or $m_u$ as the basis for a definition of the kilogram and to consider the impact of new experimental data becoming available which may resolve discrepancies in the existing set. This has also enabled us to derive, for the first time, a continuity relationship between the present values and uncertainties of $h$, $e$ and $m_u$ and their corresponding values and uncertainties subject to any valid alternative definitions of the kilogram and ampere.

In addition, we have introduced observational equations to represent the results of experiments in electrical quantum metrology aimed at determining the "exactness" of the "quantum metrology triangle" [6,7]. These enable us to test the effectiveness of these experiments in confirming the exactness of the quantum Hall and Josephson effects.

## Selection and reorganisation of observational equations

The best available values for the fundamental constants of physics are achieved by a least squares adjustment of a set of more than one hundred so-called "observational equations" [8]. These observational equations link experimentally determined data to the values of a set of "adjusted" constants. Since the experimental data include uncertainties due to experimental error and the set of equations is over determined, a weighted least-squares method is used to determine the set of adjusted constants that minimise the sum of the squared residual errors.

The work presented here involves a reduced set of the complete set of equations listed in TABLE XL of ref [8] chosen because they have the greatest influence on the adjusted values of $h$ and $e$. They are listed in Table 1 in the form used in [8]. The relative uncertainties of the experimental data used with the reduced set are shown in Figure 1. In the following paragraphs we explain why these equations have been selected for this work and how they have been re-organised.

*The role of the fine structure ($\alpha$) and Rydberg constants ($R_\infty$)*

The starting point for our analysis is the observation that the relative uncertainties of the fine-structure constant ($\alpha$) and the ratio of the Rydberg constant to the atomic weight of the electron ($R_\infty/A_r(e)$) [1] are both close to $10^{-9}$, whilst the relative uncertainties of the other data used as inputs to this work (Figure 1) are greater than $10^{-8}$. Therefore, we do not expect the values of these two constants to be influenced strongly by the other data in the adjustment calculation, so we choose to consider them as being determined directly from the appropriate experimental data and do not include them as variables in the adjustment. This choice has a number of consequences for the work presented here.

Firstly, the equations that define the Rydberg constant (combined with the atomic weight of the electron) and the fine-structure constant are not used as observational equations by [8]. In this work, they are re-organised (lines 7 and 8 of Table 1) and included in the reduced set of observational equations. Since $\alpha$ and the ratio $R_\infty/A_r(e)$ are not adjusted in this work, $e$ and $m_u$ become the adjusted constants in these observational equations. A further change resulting from not adjusting $\alpha$ and $R_\infty$ is that the equations for $R_K$ and $K_J$ (lines 3 and 5 of Table 1) can be written in their conventional forms without explicit reference to $\alpha$.

*Introduction of the atomic mass constant*

A further difference between this work and that reported in [8] is that the equations for the determination of the Faraday constant and the molar volume of silicon (lines 1 and 2 of Table 1) are re-organised to show the atomic mass constant ($m_u$) explicitly. This is possible because, as described above, $\alpha$ and $R_\infty$ are not adjusted, and therefore their ratio can be eliminated using an expression that links them to the atomic mass constant:

---

[1] We use the relative uncertainty for the experimental value of $A_r(e)$ rather than that of the adjusted value. This choice has no significant influence on the result of the work.

$$\frac{R_\infty}{A_r(e)\alpha^2} = \frac{m_u c}{2h} \tag{7}$$

For convenience, we further reduce the number of adjusted constants by determining the lattice constant of silicon ($d_{220}$) from the appropriate experimental data and combining it with $V_m$(Si) (line 1 of Table 1). Reviewing Table 1, shows that when the equations in lines 1 and 2 are written in this way, they describe two different methods for the determination of the atomic mass constant ($m_u$); one based on measurement of the molar volume of silicon and the other the determination of the Faraday constant by silver coulometry. Since the experimental data for these are independent, we combine them into a single datum.

The equations in lines 6 and 7 (Table 1) both lead to values for the ratio $h/m_u$ . The first of these arises from experimental measurements of the Rydberg constant. The second is the combination of the results from the recoil frequency of Cs atoms and the recoil velocity of Rb atoms. These two values have relative uncertainties of $9.8\times10^{-8}$ and $1.6\times10^{-8}$, and are consistent within these estimated uncertainties so they are combined here[2] to give a single datum in the adjustment.

*Introduction of $\Lambda$ and $\Delta$*

The least squares adjustment reported here is based on the equations listed in Table 2. These differ from the equations in Table 1 by the introduction of two additional variables, which are needed to evaluate alternative definitions of the SI ampere and kilogram. In the case of the kilogram, we introduce the variable $\Lambda$, which is the proportionality constant between any alternative definition for the kilogram and its current definition [3]. It can be considered to be the reciprocal of the mass of the International Prototype Kilogram (IPK) in SI kilogram.

In the case of the ampere, we introduce for the first time the variable $\Delta$, which is the proportionality constant between the magnetic constant ($\mu_0$) and the present SI value fixed by the definition of the ampere. This enables consideration of alternative definitions of the ampere by allowing $\Delta$ to "float" and where necessary to have some uncertainty whilst retaining $\mu_0=4\pi\times10^{-7}$ H m$^{-1}$ exactly in the calculation.

The variable $\Delta$ also has an impact on the electric constant ($\varepsilon_0$) through the relationship $\mu_0\varepsilon_0 c^2 = 1$ which defines its product with the magnetic constant. Thus $\Delta$ appears in $y_6$ (Table 2) from the direct reference to $\mu_0$ and also in $z_2$ , $z_3$ and $z_4$, which we now explain in more detail.

The experiments that contribute data for ($z_2$) in Table 2 measure electrostatic force and involve both a force measurement in terms of the kilogram and the relation between force and charge through Coulomb's law in which the scaling factor is $\varepsilon_0$. Thus both $\Lambda$ and $\Delta$ appear in ($z_2$) and the square root arises from the fact that the square of the measured voltage and hence $K_J^2$ is determined in the experiments. The determination of the gyromagnetic ratio of the proton ($\gamma_p$) by the low-field method involves Ampere's Law, but does not include a mechanical measurement. Hence $\Delta$

[2] allowance is made in the calculation for the correlation between line 7 and line 8 through the presence of $\alpha$ in both.

appears in ($z_3$) but $\Lambda$ does not. The Watt balance experiment does not involve $\mu_0$ or $\varepsilon_0$ so $\Lambda$ appears in ($z_4$) but $\Delta$ does not. Also, the determination of $R_\infty$ does not involve any measurements relative to the kilogram or ampere so neither appears in (5). Measurements of the von Klitzing constant ($R_K$) that support ($z_5$) are made by relating the quantised Hall resistance to the impedance of a calculable capacitor. This relates capacitance to length via $\varepsilon_0$ so $\Delta$ appears in ( $z_5$). Since the length is measured in terms of the SI metre, the result is a value for the ratio of $R_K$ to the impedance of free space $Z_0 = \mu_0 c$. If $R_K$ is accepted as being equal to $h/e^2$ then this ratio is exactly $1/2\alpha$ [9]. We return to this later.

We now have the six observational equations listed in Table 2 where the $z_i$ on the left-hand side of each equation indicate the numerical values arising from the combinations of experimental data used here. The "adjust to be equal to" symbols indicate that these equations are not exact because there is a "residual error" term that must be minimised in the least squares adjustment.

## Selection of experimental data

The principles of the experiments used to determine the $z_i$ for the least squares adjustment performed here are listed in Table 2. The relative uncertainties of the $z_i$ are shown in Figure 2. (In the case of $K_J^2 R_K$, the correlation of 0.14 between the two NIST values is allowed for in the calculation.)

We estimate that the most significant data that we have not included are those of measurements of $h/m_n$ (where $m_n$ is the mass of the neutron), which are a further factor of 8 times less accurate than $z_6$. Whilst it would be straightforward to extend our set of observational equations by the addition of further equations, it would not change any of the conclusions. Hence, we conclude that all of the most significant data has been included.

*Numerical Method*

The adjustment procedure involves solving the seven observational equations in Table 2 as a nonlinear least-squares problem associated with the observational equations

$$\mathbf{Z} = \mathbf{\Phi}(\mathbf{a}) + \boldsymbol{\eta} \tag{8}$$

in order to find the values of the set of constants (**a**) that best fit the experimental data (**Z**), taking into account the stated experimental uncertainties (u(**Z**)) and any correlations between them [10,11]. Different scenarios relating to which combinations of constants are treated as being exact are accounted for by using constraint matrices that simultaneously rescale the problem to avoid numerical difficulties associated with the vast range of numerical values. Uncertainties associated with experimental values are propagated through to those associated with the fitted parameters and related quantities (such as the residuals). The solution approach also enables us to apply constraints in such a manner that the value of a limited number of unconstrained parameters at the solution of the optimisation problem can be pre-assigned.

We present the results in terms of the residual deviations of the adjusted values from the data ( $\mathbf{Z} - \mathbf{\Phi}(\hat{\mathbf{a}})$ ) expressed in terms of the normalised residuals $r_i/u(r_i)$ where

$$r_i = (z_i - \phi_i(\hat{\mathbf{a}})) \quad (9)$$

In the scenarios we discuss here, we do not inflate the reported experimental uncertainties of data or groups of data in order to cover discrepancies between experimental data. Whilst this process is an important part of the work reported in [8], it addresses issues not directly relevant to the comparison of different scenarios considered here.

## Results of the least-squares adjustment

In the following sections, several cases are considered, according to which of the quantities are adjusted and which are fixed. These are summarised in Table 3.

*The "Base case"*

The "base case" for this work involves the implementation of constraints that represent the current SI definitions of the kilogram and ampere. These are that both $\Lambda$ and $\Delta$ are unity with no uncertainty. The results of this adjustment are compared in Table 4 with the current accepted values reported in [8].

The difference between the adjusted values of $h$, $e$ and $m_u$ in the base case and the current accepted values [8] are approximately 5 times smaller than their relative uncertainties. The relative uncertainties are principally set by the uncertainty of $z_4$ and are typically a factor of 2 smaller than those for the same quantities in [8]. This is because we have not expanded the uncertainties of any of the experimental data to give a better-balanced set of residuals [8]. This is evident in the normalised residuals shown in Figure 3, which are larger than their standard uncertainties for the experiments that determine $z_1$, $z_3$, $z_4$ and $z_5$.

The results of the base case confirm that the reduced set of observational equations and the experimental data used here capture the relevant information provided by the complete set used by [8].

*Alternative definitions for the kilogram and ampere*

We now proceed to compare two alternative definitions for the kilogram. In both cases we fix $e$ as part of a proposed new definition of the ampere, and represent alternative definitions for the kilogram by reference to a fixed value for either $h$ or $m_u$ (Table 3). The uncertainties in the cases discussed in this section are displayed in Table 4.

Fixing $h$ (and $e$) leads to the uncertainty in the value of $\Delta$ being dominated by the uncertainty in the experimental value of $z_7$. For comparison, when fixing $m_u$ (and $e$), the uncertainty in $h$ is dominated by the uncertainty in $z_6$, and $\Delta$ is dominated by the combined uncertainty of $z_6$ and $z_7$. The choice between fixing $h$ or fixing $m_u$ to define the kilogram has no effect on the uncertainty in the value of $\Lambda$, which is dominated in both cases by the uncertainty in $z_4$. Whichever of $h$ or $m_u$ is adjusted picks up the uncertainty in $h/m_u$ from $z_6$. We discuss these relationships further in the following section.

It is interesting to note that throughout the process of applying alternative constraints to the uncertainties of $\Lambda$ and $\Delta$, the values of all of the variables involved remain unchanged. Hence there would be no difficulty in maintaining the continuity of values with the present values as represented by the base case if any new definition were to be adopted. For example, the only effect on mass metrology, which currently depends on traceability to the IPK, would be to add a small uncertainty (largely due to $z_4$) to $\Lambda$. There would be no change in the value of the IPK itself when expressed in units of a re-defined kilogram.

Figure 3 shows that the largest normalised residuals that are not consistent within their standard uncertainties are those associated with $z_1$ and $z_4$. The residual associated with $z_5$ also does not cover zero, but it is significantly smaller in absolute terms. It is possible that experimental work to re-determine the Avogadro constant by the XRCD method may lead to a value that resolves the 1.2 ppm discrepancy between it and the accepted value of the Planck constant. We have modelled this possibility and the results are also shown in Figure 3. Our results suggest that the value of 1.2ppm that resolves the existing discrepancy does not reduce the residual of $z_1$ to zero. This is achieved by a correction of approximately 0.6 ppm, which resolves the residual deviation of $z_1$ completely. Nevertheless, the reconciliation of the XRCD and WB results does not resolve the discrepancy with $z_3$. This is because it is not an experiment that links mechanical to electrical units – it has no reference to the kilogram (it does not involve $\Lambda$). Although the value used here for $z_3$ arises from a single experiment, it is consistent to 1 part in $10^8$ with another experimental determination of the same quantity.

## Coherence of the observational equations

In order to gain further insight into the underlying mathematical structure of the observational equations and to demonstrate their coherence[3], we give two examples of their application. These lead to an expression that demonstrates the continuity of the current values of the variables with their values according to any valid re-definition. We start by clarifying the definitions of

$$\Lambda = \frac{m'}{m} \tag{10}$$

and

$$\Delta = \frac{\mu_0'}{\mu_0} \tag{11}$$

where $\mu_0$ and $m$ represent the values of the magnetic constant and a mass in SI units. The dashed forms represent the values of the same quantities expressed in terms of alternative definitions of the ampere and kilogram.

In the first example, we equate the force between two elementary charges according to Coulomb's law with the force on a mass under gravity

$$\frac{\mu_0 c^2}{4\pi} \frac{e'^2 \Delta}{r} = m' g \tag{12}$$

[3] the term "coherence" is used in the specialised sense in which it is used in the SI brochure, that when coherent units are used, conversion factors between units are never required.

where we have used $\Lambda$ and $\Delta$ as described above and $e'$ is the elementary charge in the re-defined units. The corresponding expression in SI units is

$$\frac{\mu_0 c^2}{4\pi}\frac{e^2}{r} = mg \qquad (13)$$

Dividing (12) by (13) with the substitution of (10) gives

$$\frac{e'^2 \Delta}{e^2} = \Lambda \qquad (14)$$

This expression describes the relationship between the values of $\Lambda$, $\Delta$ and $e'$ in any valid unit system. There are many alternative ways to derive this expression such as setting the value of the fine structure constant equal when calculated from quantities expressed in alternative units (shown with primes) as follows

$$\alpha = \frac{\mu_0 c}{2}\frac{e'^2 \Delta}{h'} = \frac{\mu_0 c}{2}\frac{e^2}{h} \qquad (15)$$

Although the same expression can be derived by equating the impedence of a Thompson Lampard calculable capacitor with that of a quantum Hall effect device, as emphasised by the derivation presented here, it does not require any reference to the quantum electrical effects, but is determined by the description of the hydrogen atom. It confirms that the relationship between the kilogram and the Planck constant has some significance that goes beyond the quantum Hall effect, the Josephson effect or even the Watt balance experiment.

*Continuity equations and uncertainties*

The continuity equation (14) together with (17), which is derived from it, provides a powerful method for calculating the uncertainty of the fundamental constants involved in any valid redefined unit system with respect to their uncertainties in SI units. In the case of a definition of the kilogram based on a fixed value of $h$ (*ie* $u(h')=0$), we can show that $u(\Lambda)=u(h)$ and $u(\Delta)=u(e^2/h)$, whilst the uncertainty in the two electrical quantum constants ($R_K$ and $K_J$) is zero. In the case of a definition of the kilogram based on a fixed value of $m_u$ (*ie* $u(m'_u)=0$), then $u(\Lambda)=u(h.m_u/h)$ and $u(\Delta)=u(e^2/h.h/m_u)$, whilst the uncertainty in the two electrical quantum constants ($R_K$ and $K_J$) is $u(h/m_u)$. It is interesting to observe that the difference between the uncertainties in these two cases is solely due to the involvement of $u(h/m_u)$ in the second case.

These results confirm the results of the least-squares adjustment presented above, where it should be noted that the correlation between the uncertainties must be taken into account when calculating numerical values (*eg* between $u(h)$ and $u(m_u/h)$).

## Adjustments including consideration of the “Quantum Metrology Triangle”

The discussion above presumes the exactness of the quantum Hall and Josephson effects as used in the observational equations for $z_2$, $z_3$ , $z_4$ and $z_5$. (The same assumptions are also central to the least-squares adjustment reported in [8].) We now introduce the variables $\varepsilon_J$ and $\varepsilon_K$ that represent possible corrections to each effect, so that these equations are replaced by

$$z_2^* \doteq \frac{2\Lambda^{1/2}e\Delta^{1/2}}{h}(1+\varepsilon_J) \qquad (16)$$

$$z_3^* \doteq \frac{m_u}{e^2\Delta}(1+\varepsilon_J)(1+\varepsilon_K) \qquad (17)$$

$$z_4^* \doteq \frac{4\Lambda}{h}(1+\varepsilon_J)^2(1+\varepsilon_K) \qquad (18)$$

$$z_5^* \doteq \frac{h}{e^2\Delta}(1+\varepsilon_K) \qquad (19)$$

These observational equations enable the two corrections to be incorporated into the adjustment in full. An evaluation of these corrections has been discussed previously [8] based on an analysis of whether there was sufficient justification in the experimental data to confirm whether they were non-zero. We use an alternative approach of incorporating them into the adjustment of the observational equations and data. This approach enables us to calculate the uncertainties in the adjusted values of $\varepsilon_J$ and $\varepsilon_K$ for various possible uncertainties in the input data. {

A strong interest in confirming the exactness of the quantum Hall effect and the Josephson effects has led to the design and execution of specialised experiments often referred to as attempts to “close the quantum metrology triangle” [6,7,12]. In order to discuss the potential impact of these experiments in greater detail, we present an adjustment that includes an additional observational equation. This represents the results of the NIST electron counting capacitance standard [13]

$$z_8 \doteq \frac{2e^2\Delta}{h}(1+\varepsilon_J) \qquad (20)$$

This experiment involves a combination of charging a capacitance with a known number of electrons and then measuring it by reference to a Thompson-Lampard capacitor. It has a published uncertainty $u_r(z_8)$ of $9.2\times10^{-7}$.

Planned experiments to apply Ohm's Law directly to simultaneous measurements of single-electron tunnelling, the quantum Hall effect and the Josephson effects give rise to an observational equation that has the same form as $z_3^*$However, since it has been suggested that it may achieve a relative uncertainty that is better than that of $z_3^*$, we also include a trial value of $u_r(z_8)= 10^{-8}$. The results of a least-squares adjustment including these additional equations are shown in Table 5.

The introduction of the additional observational equation and the two additional adjusted parameters ($\varepsilon_J$ and $\varepsilon_K$) changes the relative influence of the data on the final adjusted results.

The results of an adjustment based on present experimental data suggest that the introduction of experiments intended to "close the quantum metrology triangle" [6,7,13,14] confirm that $\varepsilon_K$ is very much smaller than $\varepsilon_J$. This is consistent with recent theoretical work that suggests that $\varepsilon_K$ may be as small as $10^{-18}$ [15]. However, the data is inconsistent because of the discrepancy between the experimental value of $z_1$, which is the only means of determining $h$ directly, without reference to $z_4$, which determines $h$ more accurately but relies on the assumptions underlying the quantum Hall and Josephson effects.

In summary, the results in Table 5 show that when $z_8$ is inaccurate (*ie* it makes very little contribution to the fit), the uncertainties of $h$, $e$ and $m_u$ are determined by $u_r(z_1)$. As $z_8$ becomes more accurate, their uncertainties become determined by a combination of $u_r(z_4)$ and $u_r(z_5)$. This suggests that a relative uncertainty of at least $10^{-8}$ for $z_8$ is likely to be required to provide useful justification for the exactness of the quantum effects driven by $u_r(z_5)$ and $u_r(z_1)$.

In addition to introducing new experimental data, it is also possible to introduce a third correction term $\varepsilon_S$ defined by $Q_S=(1+\varepsilon_S)e$ which tests the hypothesis that there is a departure in the charge observed in single-electron tunnelling experiments from the elementary charge and appears only in the expression $y_8$. The rationale for this hypothesis has been discussed elsewhere [16,17]. The effect of introducing $\varepsilon_S$ is to increase the significance of $z_1$ further and thereby to exacerbate the difficulty of its disagreement with $z_4$ leading to its uncertainty dominating the adjustment. Hence we draw the interesting conclusion that the most significant experiment bearing on the "closure of the quantum metrology triangle" concerns the determination of $z_1$, whether it is assumed that there is any uncertainty in the exactness of the elementary charge or not.

## Conclusions

We have presented a reduced set of observational equations involving a subset of fundamental constants that can be used to evaluate the influence of candidate re-definitions for the kilogram and ampere. A comparison of our base case, which corresponds to the present SI, with the accepted values of $h$ and $m_u$ confirms that the equations selected for this work capture sufficient of the observational data and the underlying physics for this type of analysis.

We have shown, for the first time, the importance of introducing the variable $\Delta$ into the observational equations in order to carry out calculations involving alternative definitions for the ampere. This has enabled us to derive a simple continuity equation

(14) linking $\Lambda$, $\Delta$ and $e$. The continuity equation shows that the continuity of the values of $e$ and $h$ between their current SI values and the values in an alternative unit system is assured. The derivation of this equation confirms that the link between the kilogram, as represented here by $\Lambda$, and the values of the Planck and atomic mass constants is independent of the quantum electrical effects.

In all of the scenarios analysed here, the areas where the data is least consistent concern the reconciliation of the results of the Si-XRCD experiment ($z_1$) with the Watt balances ($z_4$) and the determination of the gyromagnetic ratio of the proton ($\gamma_p$) by the low-field method ($z_3$). There is also a lack of consistency between the determination of the fine structure constant from the quantum Hall effect ($z_5$) and the electron magnetic moment ($z_7$). We have analysed scenarios of fixing either $h$ or $m_u$ as part of a revised definition of the kilogram. It makes little difference which of these is fixed, because their ratio is known with very good accuracy. The difference in their uncertainties is due to the inclusion of extra uncertainty due to $u_r(h/m_u)$ in the case where $m_u$ is fixed. Since this relative uncertainty is currently $1.4\times10^{-9}$ it has little practical significance.

We have considered a scenario in which a future determination of the Avogadro constant by the Si-XRCD method gives rise to data that is displaced from the present value by 1.2 ppm. Our calculations suggest that a displacement of 0.6 ppm resolves the residual deviations, but in turn only serves to emphasise the remaining discrepancy with $z_3$. This has not received much discussion, but deserves further consideration because it relates the Planck constant and the elementary charge to the electrical units by an independent experiment.

Our analysis of experiments to "complete the quantum metrology triangle" suggests that they can only confirm the exactness of the quantum physics involved when it is also assumed that the SET is exact (*ie* $\varepsilon_S=0$). Additionally, QMT experiments can provide little information about the exactness of the quantum electrical effects unless their relative uncertainty is less than $10^{-8}$ and also that the relative uncertainty of $z_1$ is $10^{-8}$ and is consistent with $z_4$ at the same level.

We observe that the pattern of residual deviations (and uncertainties) for the observational equations presented here remains the same in all of the scenarios considered. None of them can resolve discrepant data when it is assumed that the quantum electrical effects are exact. We have shown that $u_r(\Lambda)$ is driven by $u_r(h)$, which in turn should be expanded from the value of $3\times10^{-8}$ used here to $5\times10^{-8}$ [8] to accommodate the discrepancies within the data. Since this is more than the hypothesised drift in the mass of the IPK over several decades [17] we suggest that the data is not sufficiently self-consistent to support a proposal for re-definition at present.

*Acknowledgements*

The support of Thomas Gillam (Oxford University) in carrying out some of the computations and preparation of the diagrams is gratefully acknowledged.

## References

1. BIPM 2006 *The International System of Units* 8th edn (Sevres, France; Bureau International des Poids et Mesures).
2. B.N. Taylor and P.J. Mohr, "On the redefinition of the kilogram" *Metrologia* vol 36 pp 63-64, 1999.
3. Mills I M, Mohr P J, Quinn T J, Taylor B N and Williams E R, "Redefinition of the kilogram: a decision whose time has come" *Metrologia*, vol 42 pp 71-80, 2005.
4. Becker P, De Bievre P, Fujii K, Glaeser M, Inglis B, Luebbig H and Mana G, "Considerations on future redefinitions of the kilogram, the mole and of other units" *Metrologia*, vol 44, pp 1-14, 2007.
5. Milton M J T, Williams J M and Bennett S J, "Modernising the SI: towards an improved, accessible and enduring system" *Metrologia*, vol 44, pp 356-364, 2007.
6. J Flowers, "The route to atomic and quantum standards" *Science* vol 306, pp1324-1330, 2004.
7. Gallop JC, "The quantum electrical triangle" *Phil Trans Roy Soc*, vol 363, pp2221-2247, 2005
8. P. J. Mohr and B. N. Taylor, "CODATA recommended values of the fundamental constants: 2006" *Rev Mod Phys*, vol 80, pp 633-730, 2008.
9. K. von Klitzing, G. Dorda and M. Pepper, "New method for high-accuracy determination of the fine-structure constant based on quantised Hall resistance" *Phys. Rev. Lett.* **45 pp** 494-497, 1980.
10. M. G. Cox, A. B. Forbes, J. Flowers and P. M. Harris, "Least squares adjustment in the presence of discrepant data" *Advanced Mathematical and Computational Tools in Metrology VI,* (World Scientific, Singapore) pp37-51, 2004.
11. P. J. Mohr and B. N. Taylor, "CODATA recommended values of the fundamental constants: 1998" *Rev Mod Phys*, vol 72, pp 351-495, 2000.
12. A.A. Penin, *Phys. Rev. B.* **79** 113303, 2009.
13. M. Keller, "Uncertainty budget for the NIST electron counting capacitance standard, ECCS-1" *Metrologia*, vol 44, pp 505-512, 2007.
14. M. Keller, "Current status of the quantum metrology triangle" *Metrologia*, vol 45, pp 102-109, 2008.
15. F. Piquemal and G. Geneves, "Argument for a direct realization of the quantum metrological triangle" *Metrologia* vol 37 pp 207-211, 2000.
16. M Stock and TJ Witt, "CPEM 2006 round table discussion – proposed changes to the SI" *Metrologia*, vol 43, pp 583-587, 2006.
17. M.W. Keller, F. Piquemal, N. Feltin, B. Steck and L. Devoille, "Metrology triangle using a Watt balance, a calculable capacitor and a single-electron tunnelling device" *Metrologia* vol 45 pp 330-334, 2008.
18. R.S. Davis *Phil. Trans. R. Soc. London* A **363** pp 2249-64 2005.

| Line | Observational equation in form used in reference[8] | Observational equation in reorganised form | $z_i$ |
|---|---|---|---|
| 1 | $V_m(Si) \doteq \sqrt{2} M_u c \dfrac{A_r(e) d_{220}^3}{R_\infty} \dfrac{\alpha^2}{h}$ | $V_m(Si)/d_{220}^3 \doteq 2\sqrt{2} \dfrac{M_u}{m_u}$ | $z_1$ |
| 2 | $F_{90} \doteq \dfrac{cM_u}{K_{J-90} R_{K-90}} \dfrac{A_r(e)}{R_\infty} \dfrac{\alpha^2}{h}$ | $F_{90} \doteq \dfrac{2}{K_{J-90} R_{K-90}} \dfrac{M_u}{m_u}$ | |
| 3 | $K_J \doteq \left( \dfrac{8\alpha}{\mu_0 c h} \right)^{1/2}$ | $K_J \doteq \dfrac{2e}{h}$ | $z_2$ |
| 4 | $\boldsymbol{\Gamma'_p = \dfrac{K_{J90} R_{K90}}{2\mu_0} \left( \dfrac{\mu'_p}{\mu_e} \right) \dfrac{g_e \alpha^3}{R_\infty}}$ | $\boldsymbol{\dfrac{\Gamma'_p R_\infty}{c \alpha^2 g_e} \left( \dfrac{\mu_e}{\mu'_p} \right) = \dfrac{K_{J-90} R_{K-90}}{4} \dfrac{e^2}{h}}$ | $z_3$ |
| 4 | $K_J^2 R_K \doteq \dfrac{4}{h}$ | $K_J^2 R_K \doteq \dfrac{4}{h}$ | $z_4$ |
| 5 | $\boldsymbol{R_K \doteq \dfrac{\mu_0 c}{2\alpha}}$ | $R_K \doteq \dfrac{h}{e^2}$ | $z_5$ |
| 6 | $\dfrac{h}{m(\mathrm{X})} \doteq \dfrac{c}{2} \dfrac{A_r(\mathrm{e})}{A_r(\mathrm{X}) R_\infty} \alpha^2$ | $\dfrac{h}{m(\mathrm{X})} A_r(\mathrm{X}) \doteq \dfrac{h}{m_u}$ | $z_6$ |
| 7 | $m_u = \dfrac{2R_\infty}{A_r(e)} \dfrac{h}{\alpha^2}$ | $\dfrac{A_r(e)}{R_\infty} \alpha^2 \doteq \dfrac{2}{c} \dfrac{h}{m_u}$ | |
| 8 | $e = \left( \dfrac{2\alpha h}{\mu_0 c} \right)^{1/2}$ | $\alpha^{-1} \doteq \dfrac{2h}{\mu_0 c e^2}$ | $z_7$ |

**Table 1** Summary of observational equations and their re-organisation for this work. The equation in line 6 is applied to X=$^{133}$Cs and $^{89}$Rb. The equations on lines 7 and 8 are not observational equations in ref [8], but are re-organised to be observational equations here. The final column indicates which of the equations in Table 2 has been formed from each line.

| Equation | Number | Input data |
|---|---|---|
| $z_1 \doteq \frac{\Lambda}{m_u}$ | (1) | Two values of $m_{u}$. One from the molar volume of silicon and one from the Faraday constant |
| $z_2 \doteq \frac{2\Lambda^{1/2} e \Delta^{1/2}}{h}$ | (2) | Two values of $K_J$ from the determination of the Josephson constant with electrostatic "voltage" balances. |
| $z_3 = \frac{m_u}{e^2 \Delta}$ | (3) | One value from an experimental determination of the gyromagnetic ratio of the proton by the low-field method. |
| $z_4 \doteq \frac{4\Lambda}{h}$ | (4) | Three values of $K_J^2 R_K$ determined using Watt balances. |
| $z_5 \doteq \frac{h}{e^2 \Delta}$ | (5) | Five values of $R_K$ determined using Thompson-Lampard calculable capacitors. |
| $z_6 \doteq \frac{h}{m_u}$ | (6) | Three values for $h/m_u$. Two from direct measurements of $^{133}$Cs and $^{87}$Rb, and one calculated from the CODATA value [8] of the Rydberg constant |
| $z_7 \doteq \frac{h}{e^2 \Delta}$ | (7) | Two experimentally determined values of $\alpha$. |

**Table 2**: The six observational equations adjusted in this work (in order of decreasing relative uncertainty). The equation numbers are referred to in the text. The sources of the experimental data used to form the $y_i$ are listed in the final column.

| Case | Fixed quantities | Adjusted quantities |
|---|---|---|
| "base case" | $\Delta$ and $\Lambda$ | $e$, $h$ and $m_u$ |
| ampere defined by $e$ and kilogram defined by $h$ | $e$ and $h$ | $\Delta$, $\Lambda$ and $m_u$ |
| ampere defined by $e$ and kilogram defined by $m_u$ | $e$ and $m_u$ | $\Delta$, $\Lambda$ and $h$ |

**Table 3** Summary of fixed and adjusted values in the cases considered here.

| | base case | | fixed $e$ and $h$ | fixed $e$ and $m_u$ |
|---|---|---|---|---|
| $a$ | $(a-a_0)/a_0$ | $u_r(a)$ | $u_r(a)$ | $u_r(a)$ |
| $e$ | -0.310 | 1.563 | 0.000 | 0.142 |
| $h$ | -0.632 | 3.125 | 0.000 | 0.000 |
| $m_u$ | -0.650 | 3.128 | 0.142 | 0.000 |
| $\Lambda$ | 0.000 | 0.000 | 3.125 | 3.129 |
| $\Delta$ | 0.000 | 0.000 | 0.068 | 0.208 |

**Table 4** Comparison of adjusted values (***a***) with respect to the results of CODATA 06 ($\boldsymbol{a_0}$). The relative uncertainties are shown for the base case and two alternatives (all $\times 10^8$)

| | $u_r(z_7) = 1.1 * 10^{-7}$ $u_r(z_8) = 9.2 * 10^{-7}$ | | $z1 + 0.6$ *ppm* | | $z1 + 1.2$ *ppm* | |
|---|---|---|---|---|---|---|
| $a$ | $(a-a_0)/a_0$ | $u_r(a)$ | $(a-a_0)/a_0$ | $u_r(a)$ | $(a-a_0)/a_0$ | $u_r(a)$ |
| $e$ | -8.2 | 9.5 | 0.69 | 9.5 | 9.6 | 9.5 |
| $h$ | -16 | 19 | 1.4 | 19 | 19 | 19 |
| $m_u$ | -16 | 19 | 1.4 | 19 | 19 | 19 |
| $1+\varepsilon_J$ | -7.9 | 9.4 | 0.8 | 9.4 | 9.5 | 9.4 |
| $1+\varepsilon_K$ | 1.0 | 0.0 | 1.0 | 0.0 | 1.0 | 0.0 |

**Table 5** Results of adjustment of the nine-equation model including the correction terms for non-exact quantum electrical effects (all $\times 10^8$). For each of these adjustments $\Lambda$ , $\Delta$ and $1+\varepsilon_K$ have been set equal to unity.

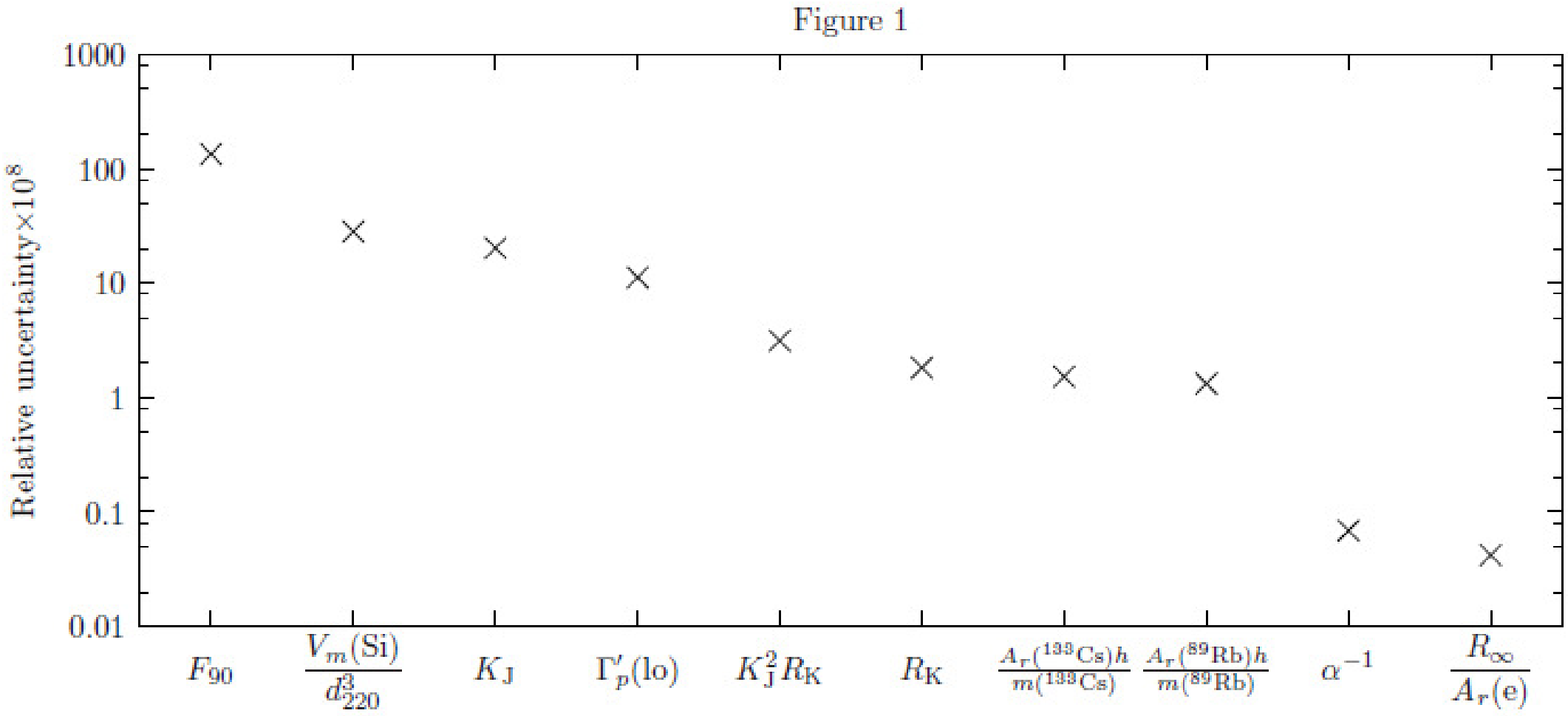

**Figure 1**: The relative uncertainties of the experimental data used as inputs to the observational equations listed in Table 2.

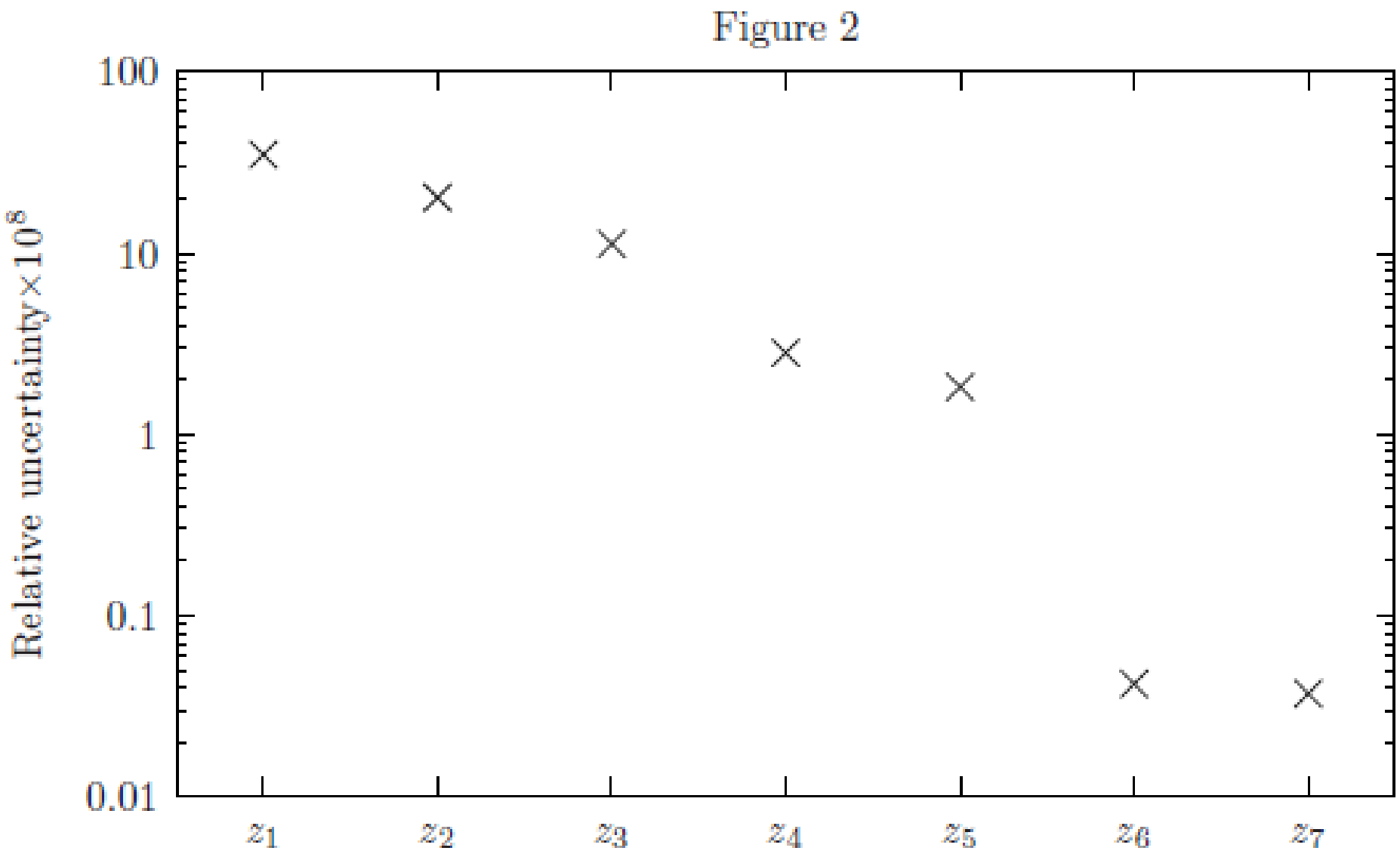

**Figure 2**: The relative uncertainties of the $z_i$ used in the observational equations listed in Table 2.

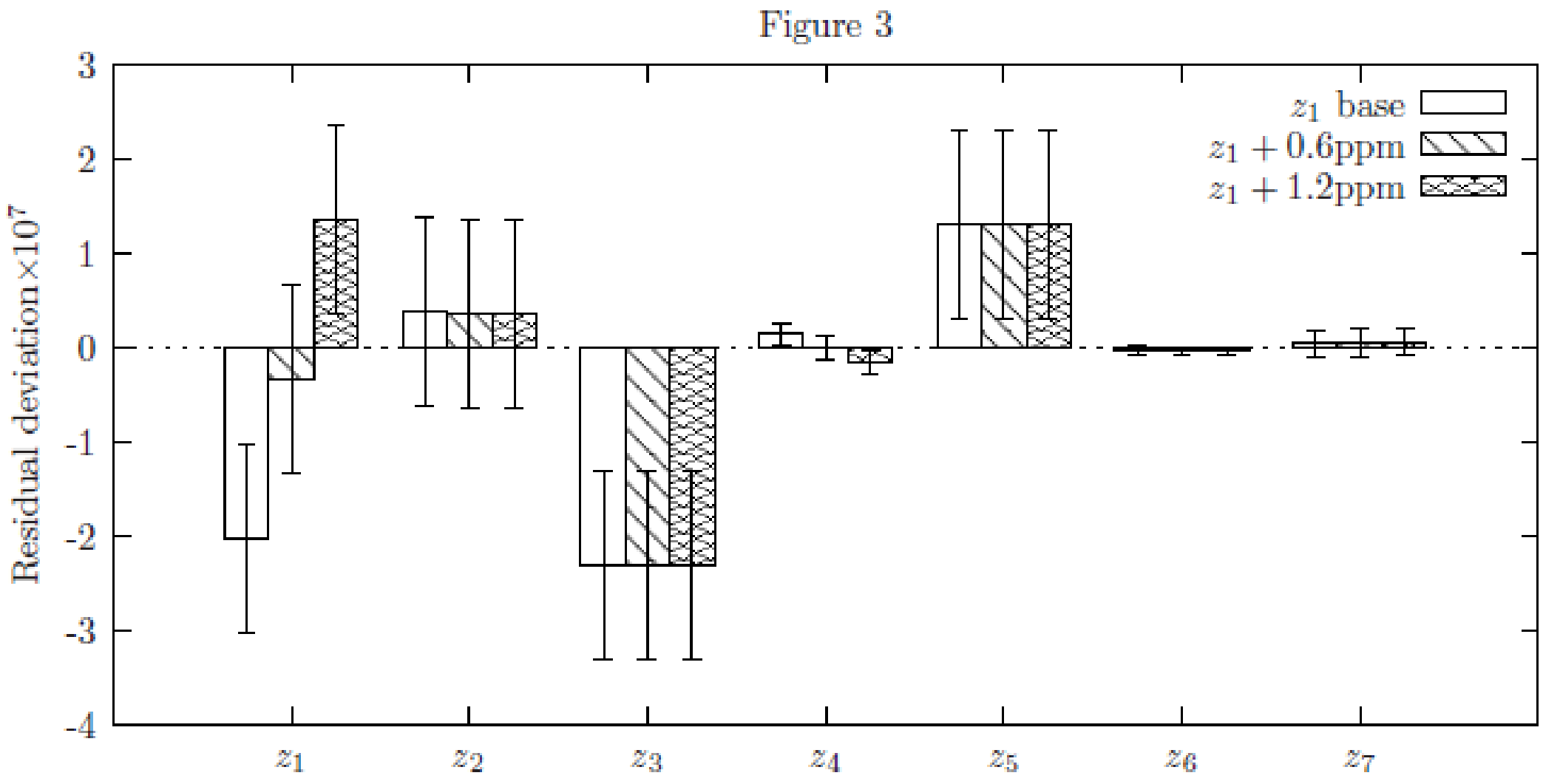

**Figure 3:** Normalised residual deviations $\times 10^8$ for the seven adjusted observational equations in the "base case". The bars indicate the standard uncertainty. The residual deviations for cases where $z_1$ is displaced by 0.6 and 1.2 ppm are also shown.